\documentstyle[psfig,float]{mn}

\title[Axisymmetric Modes of Rotating Relativistic Stars]
      {Axisymmetric Modes of Rotating Relativistic Stars in the Cowling
       Approximation}
\author[Jos{\'e} A. Font,
        Harald Dimmelmeier,
        Anshu Gupta and
        Nikolaos Stergioulas]
       {J.A. Font$^1$,
        H. Dimmelmeier$^1$,
        A. Gupta$^2$ and
        N. Stergioulas$^3$ \\
        $^1$Max-Planck-Institut f{\"u}r Astrophysik,
        Karl-Schwarzschild-Str. 1, D-85741, Garching, Germany \\
        $^2$Raman Research Institute,
        C.V. Raman Avenue, Sadashivanagar, Bangalore 560080, India \\
        $^3$Department of Physics,
        Aristotle University of Thessaloniki,
        Thessaloniki 54006, Greece}

\date{}
\pagerange{}
\pubyear{}

\begin{document}

\maketitle

\label{firstpage}

\begin{abstract}
  Axisymmetric pulsations of rotating neutron stars can be excited in
  several scenarios, such as core-collapse, crust and core-quakes and
  binary mergers and could become detectable either in gravitational
  waves or high-energy radiation. Here, we present a comprehensive
  study of all low-order axisymmetric modes of uniformly and
  rapidly rotating relativistic stars. Initial stationary
  configurations are appropriately perturbed and are numerically
  evolved using an axisymmetric, nonlinear relativistic hydrodynamics
  code, assuming time-independence of the gravitational field (Cowling
  approximation). The simulations are performed using a
  high-resolution shock-capturing finite-difference scheme accurate
  enough to maintain the initial rotation law for a large number of
  rotational periods, even for stars at the mass-shedding limit.
  Through Fourier transforms of the time evolution of selected fluid
  variables, we compute the frequencies of quasi-radial and non-radial
  modes with spherical harmonic indices $l=0,1,2$ and $3$, for a
  sequence of rotating stars from the non-rotating limit to the
  mass-shedding limit. The frequencies of the axisymmetric modes are
  affected significantly by rotation only when the rotation rate
  exceeds about 50\% of the maximum allowed. As expected, at large
  rotation rates, apparent mode crossings between different modes
  appear. In addition to the above modes, several axisymmetric
  inertial modes are also excited in our numerical evolutions.
\end{abstract}


\section{Introduction}                                                  %
\label{intro}                                                           %

The pulsations of rotating neutron stars are expected to be a source
of detectable gravitational waves. Additionally, their excitation
could become detectable in the emission of high-energy radiation.  In
particular, axisymmetric oscillations can be excited in a number of
different astrophysical scenarios, namely: a) after a core-collapse
leading to a supernova explosion (see e.g. M{\"o}nchmeyer et al 1991;
Zwerger \& M{\"u}ller 1997), b) during starquakes induced by the secular
spin-down of a pulsar, c) after a large thermonuclear explosion in the
crust of an accreting neutron star, d) during a core-quake due to a
large phase-transition to, for example, strange quark matter (Cheng \&
Dai 1998), and e) in the delayed collapse of the merged object in a
binary neutron star merger (Ruffert, Janka \& Sch{\"a}fer 1996; Shibata \&
Uryu 2000). The observational detection of such pulsations will yield
valuable information about the equation of state of relativistic stars
(see Kokkotas, Apostolatos \&  Andersson, 2000, see also Kokkotas and
Schmidt 1999, for a recent review on oscillations of relativistic
stars).

Numerical simulations of some of these scenarios are available and
provide very detailed information of the dynamics of the neutron star
pulsations.  In particular, the axisymmetric core-collapse simulations
of M{\"o}nchmeyer et al (1991) and Zwerger \& M{\"u}ller (1997), revealed that,
after the collapse and bounce of an iron core, the unshocked inner
core (the proto-neutron star) oscillates with various volume (radial
and quasi-radial) and surface modes. The amplitude and frequency of
these fluid modes ($f$- and $p$-modes) was found to depend on the
kinetic energy of the inner core at bounce, the stiffness of the
equation of state (EOS), and the central and average density of the
inner core. These authors found that the amplitude of the post-bounce
oscillations is small for spherical models, being strongly damped
through the emission of asymmetric pressure waves, in time scales of
the order of 1ms.  However, for rotating cores which bounce due to
centrifugal forces at subnuclear densities, much larger amplitudes are
achieved (as large as 10 times the central density) and the damping
time scale becomes comparable to the oscillation time scale ($\gg$ 1ms).
Recently, Dimmelmeier, Font \& M{\"u}ller (2000) have developed a code to
study axisymmetric core-collapse in general relativity using the
conformally flat metric approach (Wilson, Mathews \& Marronetti 1996).
This code is currently being applied to collapse some of the initial
models of Zwerger \& M{\"u}ller (1997), to analyze the gravitational waves
emitted in the process. Excitation of axisymmetric modes have already
been observed in such relativistic core-collapse simulations.

Of all axisymmetric modes, the quasi-radial modes of slowly-rotating
relativistic stars were first studied by Hartle \& Friedman (1975) and,
more recently, by Datta et al. (1998). In rapid rotation, quasi-radial
modes of relativistic stars have been studied by Yoshida \& Eriguchi
(2000) in the Cowling approximation (McDermott, Van Horn \& Scholl 1983), 
i.e., by neglecting the perturbations in the gravitational field (see
Stergioulas 1998 for a recent review on the equilibrium structure and
oscillations of rapidly rotating stars in general relativity). For
Newtonian stars, axisymmetric modes have been extensively studied by
Clement (1981, 1984, 1986). From the above studies, it has become
apparent that rotation weakly modifies the oscillation frequencies for
low-order modes, but introduces apparent crossings between
higher-order modes for rapidly rotating models. In addition, in Clement
(1981) it is claimed that the axisymmetric quadrupole $f$-mode lies on a
continuous branch for rapidly rotating Newtonian stars.

In this paper we compute all low-order $l=0$, 1, 2 and 3 axisymmetric
modes for rapidly rotating stars in general relativity, in the Cowling
approximation. For this purpose, we use a 2-D nonlinear hydrodynamics
code, whose accuracy has been extensively tested in Font, Stergioulas
\& Kokkotas (2000) (hereafter FSK; see also Stergioulas, Font \&
Kokkotas 1999). This code is based on high-resolution shock-capturing
(HRSC) finite-difference schemes for the numerical integration of the
general relativistic hydrodynamic equations (see Font 2000 for a
recent review).  We note in passing that the 3-D version of the
numerical methods employed here has been recently applied by
Stergioulas \& Font (2000) in the study of the large-amplitude,
nonlinear evolutions of $r$-modes in rotating relativistic stars. In
our present study of axisymmetric modes, we focus on a sequence of
equilibrium models with a polytropic ($N=1.0$) equation of state and
uniform rotation.  For the excitation of the various oscillation
modes, low-amplitude perturbations (using appropriate trial
eigenfunctions) are added to the initial equilibrium models. The
Cowling approximation allows us to evolve relativistic matter for a
much longer time than presently available coupled spacetime plus
hydrodynamical evolution codes (Alcubierre et al. 2000; Font et al.
2000; Shibata, Baumgarte \& Shapiro 2000; Shibata \& Uryu 2000). This is
particularly evident when hydrodynamically evolving rotating stars.
Nevertheless, since pulsations of neutron stars are mainly a
hydrodynamical process, the approximation of a time-independent
gravitational field still allows for qualitative conclusions to be
drawn, when studying the evolution of perturbed rotating neutron
stars. In addition, our present results will serve as test-beds for
3-D general-relativistic evolution codes.

The paper is organized as follows: In Section 2 we describe the setup
of the problem, by briefly presenting some details of the initial
equilibrium stellar configurations and the main features of the
hydrodynamical code. In Section 3 we explain the procedure by which
the initial equilibrium models are perturbed. Section 4 presents the
main results of our simulations, including the frequencies of all
low-order axisymmetric modes for our sample of initial models. The
paper ends with Section 5 where a summary is presented, together with
an outlook of possible future directions of this investigation.

\section{Problem Setup}                                                 %
\label{setup}                                                           %

Our initial models are numerical solutions of the exact equations
describing rapidly rotating relativistic stars, having uniform angular 
velocity $\Omega$. We assume a perfect fluid, zero-temperature equation 
of state (EOS), for which the energy density is a function of pressure only. 
The following relativistic generalization of the Newtonian polytropic EOS
is chosen:
\begin{eqnarray}
p&=&K\rho_0^{1+1/N},
\\
\epsilon&=& \rho_0 +Np,
\end{eqnarray}
\noindent
where $p$ is the pressure, $\epsilon$ is the energy density, $\rho_0$ is the
rest-mass density, $K$ is the polytropic constant and $N$ is the
polytropic exponent. The initial equilibrium models are computed using
a numerical code developed by Stergioulas \& Friedman (1995). Our
representative neutron star models are characterized by $N=1$, $K=100$
and central density $\rho_{\mbox{c}}=1.28\times 10^{-3}$, in units in which
$c=G=M_\odot=1$. We compute 12 different initial models by varying the
polar to equatorial circumferential radius from 1 (non-rotating star)
to $0.65$ (near the mass-shedding limit), as listed in Table 1. The
angular velocity at the mass-shedding limit is $\Omega_K= 0.8094\times10^4$
s$^{-1}$ for this sequence of rotating relativistic stars of same
central density.  In order to be able to study stellar pulsations the
initial model is supplemented by a uniform, non-rotating ``atmosphere''
of very low density, typically $10^{-6}$ or less times the central
density of the star (see related discussion in FSK).

\begin{table}
\centering
\begin{minipage}{65mm}
\caption{Equilibrium properties of the initial models, as
  described by a polytropic EOS, $p=K\rho_0^{1+1/N}$, where $N=1$,
  $K=100$ and with central rest-mass density $\rho_{\mbox{c}}=1.28 \times
  10^{-3}$ (in units with $c = G = M_{\odot} = 1$).  The entries in the
  table are as follows: $\Omega$ is the angular velocity of the star, $M$
  and $M_0$ are the gravitational and rest mass, $T/W$ is the ratio of
  rotational to gravitational binding energy and $R$ is the equatorial
  circumferential radius.}
\begin{tabular}{*{5}{c}}
\hline
 $\Omega $     &    $M$   &   $M_0$  &  $T/W$            & $R$  \\[0.5ex]
($10^4 s^{-1}$)& (M$_\odot$)& (M$_\odot$)&($\times 10^{-2}$) & (km) \\[0.5ex]
\hline
0.0   &  1.400 & 1.506 & 0.0   &  14.15 \\[0.5ex]
0.218 &  1.432 & 1.541 & 1.200   &  14.51 \\[0.5ex]
0.306 &  1.466 & 1.579 & 2.438 &  14.92 \\[0.5ex]
0.371 &  1.503 & 1.619 & 3.701 &  15.38 \\[0.5ex]
0.399 &  1.523 & 1.641 & 4.339 &  15.63 \\[0.5ex]
0.423 &  1.543 & 1.663 & 4.976 &  15.91 \\[0.5ex]
0.445 &  1.564 & 1.686 & 5.609 &  16.21 \\[0.5ex]
0.465 &  1.585 & 1.709 & 6.232 &  16.52 \\[0.5ex]
0.482 &  1.607 & 1.733 & 6.839 &  16.87 \\[0.5ex]
0.498 &  1.627 & 1.756 & 7.419 &  17.25 \\[0.5ex]
0.511 &  1.647 & 1.778 & 7.959 &  17.68 \\[0.5ex]
0.522 &  1.666 & 1.798 & 8.439 &  18.15 \\[0.5ex]
\hline
\label{initial_models}
\end{tabular}
\end{minipage}
\end{table}

The initial data are subsequently evolved in time with a hydrodynamics
code. The (axisymmetric) hydrodynamic equations are written as a
first-order flux-conservative system which expresses the conservation
laws of mass, momentum and energy. The specific form of the equations
was presented in FSK and we will not repeat it here. These equations
are solved using a HRSC finite-difference scheme (Font 2000). A
comprehensive description of the specific numerical techniques we
employ was previously reported in FSK. Therefore, we only mention here
that the code makes use of the third-order piecewise parabolic method,
PPM (Colella \& Woodward 1984), for the cell-reconstruction procedure,
together with Marquina's flux-formula (Donat et al 1998) to compute
the numerical fluxes. The PPM reconstruction scheme was shown to be
accurate enough for maintaining the initial rotation laws during many
rotational periods.

The hydrodynamic equations are implemented in the code using spherical
polar coordinates $(r,\theta,\phi)$ and assuming axisymmetry, i.e., all
derivatives with respect to the $\phi$ coordinate vanish.  The radial
computational domain extends to 1.2 times the radius of the star (the
20\% additional zones are used for the atmosphere). In the polar
direction, the selected domain depends on the spherical harmonic index
of the pulsation modes: for even $l$ modes the domain extends from
$\theta=0$ (pole) to $\theta=\pi/2$ (equator).  For odd $l$ modes, the domain
extends to $\theta=\pi$.  The number of grid points we employ is $200\times 80$
for $l$ being even and $160 \times 120$ for $l$ being odd, in $r$ and $\theta$
respectively. The boundary conditions are implemented in the same way
as in FSK.

Our numerical evolution code was thoroughly tested in FSK, by
comparing evolutions of perturbed spherical stars with results from
perturbation theory obtained with an independent eigenvalue code.
Since then, the code has been upgraded to run efficiently on a NEC
SX-5/3C vector supercomputer. This modification was essential for
doing a large number of numerical evolutions for many rotational periods
and with a large number of grid-points.

\section{Perturbation of the Initial Data}                              %
\label{Initial}                                                         %

The accurate computation of mode frequencies in a rotating star
requires an appropriate excitation of the equilibrium initial data.
When doing so, it is possible to obtain the frequencies of the excited 
modes with good accuracy, through a Fourier transform of the time evolution 
of the hydrodynamical variables, provided that the evolution time is much 
larger than the period of oscillations. As in the hydrodynamical evolution 
we are using the 3+1 formulation (Banyuls et al 1997), the oscillation 
frequencies of the various evolved variables are obtained with respect to 
the coordinate time at a given location. This corresponds to the frequency of 
oscillations in a reference frame attached to an inertial observer at infinity.
To increase the accuracy in the computation of the frequencies, we search 
for the zeros of the first derivative of the Fourier transform (with respect 
to the frequency), using second-order accurate central differences. These 
zeros correspond to maxima in the Fourier transform, which (except for 
high-frequency noise) correspond to the excited modes of oscillation. This 
procedure is done at several points inside the star and the frequencies thus
determined are found to be the same for each mode, i.e., all the modes
that we identify are global discrete modes. For the resolution
employed we estimate the accuracy of the computed frequencies to be of
the order of $1\%-2$\%. The different pulsation frequencies are
identified with specific normal modes by comparing frequencies of the
non-rotating star to known eigenfrequencies from perturbative
normal-mode calculations.

As demonstrated in FSK, the small-amplitude pulsations in the
nonlinear, fixed spacetime evolutions correspond to linear normal
modes of pulsation in the relativistic Cowling approximation
(McDermott et al 1983), in which perturbations of the spacetime are
ignored.  The existence of a numerical viscosity inherent to the
numerical scheme, damps the pulsations of the star. Therefore,
high-resolution grids are preferred to reduce the damping, especially
for the higher frequency modes, which are damped faster. In addition,
our numerical scheme requires the presence of a tenuous,
constant-density ``atmosphere" surrounding the star, which is reset to
its initial state after each time-step (in such a way that the stellar
surface is allowed to contract or expand), introduces an additional
numerical damping of the pulsations, due to the finite-differencing at
the surface of the star. In order to minimize this effect, the density
of the ``atmosphere" has to be small enough to be dynamically
unimportant. As already mentioned, a typical value of $10^{-6}
\rho_{\mbox{c}}$ is appropriate for this purpose.

\begin{figure}
\centerline{\psfig{file=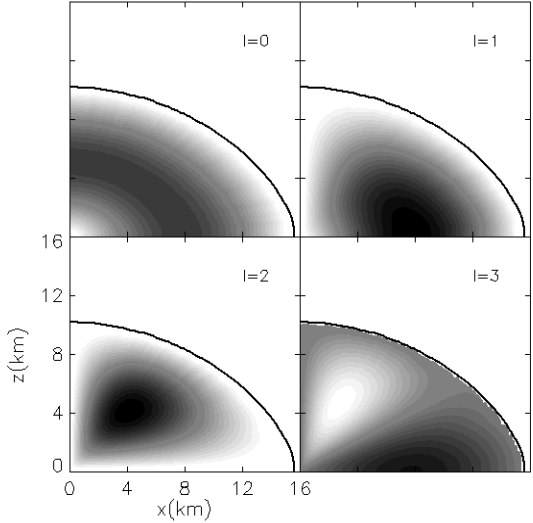,width=7.5cm,height=7.5cm}}
\caption{2-D greyscale plot of the initial data used for
  mode-excitation: density perturbation ($l=0$) and $v_{\theta}$ ($l\neq 0$).
  The darkest area corresponds to the maximum of the perturbation,
  while the lightest area corresponds to its minimum values. The thick
  solid line indicates the location of the surface of the star. The
  depicted initial model is the fastest rotator of our sample in Table
  1.}
\label{pert}
\end{figure}

We use analytic eigenfunctions to excite particular oscillation modes.
For the $l=0$ modes the initial equilibrium values of density and
pressure, $\rho_0$ and $p_0$, are perturbed to nonequilibrium values,
$\rho=\rho_0 + \delta\rho$ and $p=p_0+\delta p$ by the eigenfunctions
\begin{eqnarray}
\delta\rho &=& A \rho_c \sin\left(\frac{\pi r}{r_s(\theta)}\right),
\label{rho_pert}
\\
\delta p &=& \Gamma p_i \frac{\delta\rho}{\rho_i},
\label{p_pert}
\end{eqnarray}
where $\rho_c$ is the central density, $r_s(\theta)$ is the coordinate radius
of the surface of the star (which depends on the polar angle $\theta$) and
$\Gamma $ is the adiabatic index of the ideal gas EOS, $p=(\Gamma -1)\rho\epsilon$,
related to the polytropic index by the equation $\Gamma =1+1/N$ (for
isentropic stars). The amplitude of the excitation, $A$, is typically
chosen to be in the range 0.001 to 0.005.

For the excitation of the $l=1, 2$ and 3 modes we add a small non-zero
$\theta$-velocity component to perturb the initial vanishing value. More
precisely, for $l=1$ we have
\begin{eqnarray}
v_{\theta}=A\sin\left(\frac{\pi r}{r_s(\theta)}\right) \sin\theta,
\end{eqnarray}
for $l=2$
\begin{eqnarray}
v_{\theta}=A\sin\left(\frac{\pi r}{r_s(\theta)}\right) \sin\theta \cos\theta,
\end{eqnarray}
and for $l=3$
\begin{eqnarray}
v_{\theta}=A\sin\left(\frac{\pi r}{r_s(\theta)}\right) \sin\theta
(1-5\cos^2\theta).
\end{eqnarray}
The particular radial dependence in the above eigenfunctions is chosen
so that the perturbations vanish at the surface of the rotating star.
We have found that this is necessary for our numerical scheme to work.
Figure~\ref{pert} displays a 2-D greyscale plot of the above
perturbations, for the model near the mass-shedding limit. The darkest
area corresponds to the maximum of the perturbation, while the
lightest area corresponds to its minimum values. We note that all
models have equatorial plane symmetry with respect to the $z=0$ axis,
even though we use a grid extending to $\theta = \pi$ for odd $l$'s.  
The choice of these trial eigenfunctions allows us to compute the four
lowest frequencies quite accurately for all the considered $l$ modes,
as we discuss next.

\section{Results}                                                       %
\label{Results}                                                         %

We turn now to presenting our numerical results concerning the
frequencies of axisymmetric pulsations of uniformly and
rapidly rotating neutron stars. 

\subsection{Quasi-radial modes}

\begin{figure}
\centerline{\psfig{file=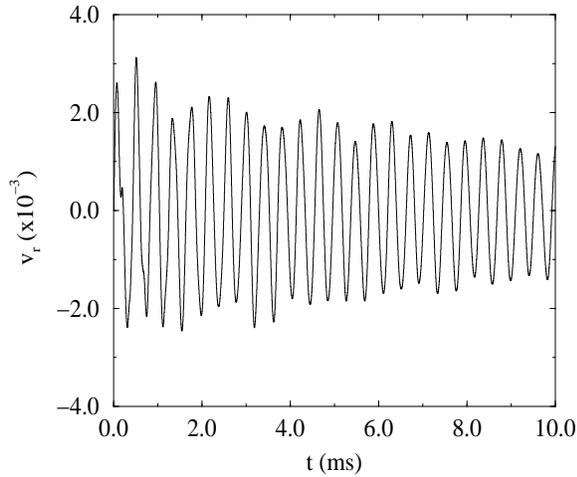,width=7.5cm,height=6.5cm}}
\caption{Time evolution of the radial velocity of the most rapidly rotating
  model of our sequence, $\Omega=0.522\times 10^4 s^{-1}$. An $l=0$ perturbation
  has been applied to the equilibrium data. The pulsations are mainly
  a superposition of the normal modes of the star.}
\label{l=0velr}
\end{figure}

Figure~\ref{l=0velr} shows the time evolution of the radial velocity
component for the most rapidly rotating model in our sequence. The
initial equilibrium model has been perturbed with an $l=0$
perturbation according to Eqs.~(\ref{rho_pert}) and (\ref{p_pert}).
The final evolution time corresponds to 10ms and the oscillations are
measured at half the radius of the star and at an angle of $\theta=\pi/4$.
The oscillatory pattern depicted in this figure is typical to all our
simulations: it is mainly a superposition of the lowest-order normal
modes of the fluid. The amplitude of the oscillations is damped due to
the inherent viscosity of the numerical scheme. The high frequency
normal modes are usually damped faster and at the final time the star 
is pulsating mostly in a few lowest frequency modes.

\begin{figure}
\centerline{\psfig{file=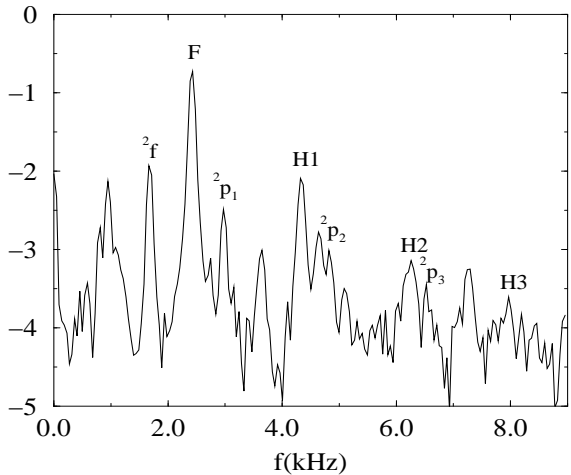,width=7.5cm,height=6.5cm}}
\caption{Logarithm of the amplitude (in arbitrary scale) resulting from the
  Fourier transform of the radial velocity evolution shown in
  Fig.~\ref{l=0velr}.  It is possible to identify in this plot the
  frequencies of the fundamental quasi-radial mode as well as up to
  three harmonics. Additionally, $f$ and overtones of $p-$modes can
  also be identified.}
\label{l=0velr_fft}
\end{figure}

The frequencies of the axisymmetric modes are obtained by a Fast
Fourier Transform of the time evolution of selected hydrodynamical
variables (both the density and the components of the velocity).
Figure~\ref{l=0velr_fft} shows the Fourier transform of the radial 
velocity evolution depicted in Fig.~\ref{l=0velr}. The main mode which 
is excited is the fundamental quasi-radial $F$-mode. Its higher harmonics 
($H1$-$H3$) are also excited, as well as several other, non-radial modes. 
The amount of excitation of each mode depends on the correlation between 
the mode eigenfunction and the applied perturbation. 

As it is apparent in Figure~\ref{l=0velr_fft}, a dense spectrum of modes
appears when one applies generic perturbations that do not
correspond to the eigenfunctions of a particular normal mode only.
Thus, in order to identify the peaks in the Fourier transform with
specific normal modes, we rely on the previously known frequencies in
the non-rotating limit (see FSK). As the rotation rate is increased, we
follow the change in the location of the various peaks, keeping in
mind that at large rotation rates apparent crossings of frequencies
occur. In such a case, the amplitude of the Fourier transform at
various points inside the star (which correlates with the mode
eigenfunction) is used as a guide in deciding about the correct
identification of the mode frequency. The specific values for
the frequencies of the fundamental $F$ and higher harmonics  
$H_1, H_2, H_3$ quasi-radial modes for the sequence of rotating 
stars considered here, are shown in Table~\ref{qr_tab}.

\begin{table}
\centering
\begin{minipage}{63mm}
\caption{Fundamental, first, second and third overtones ($F, H_1, H_2$ and $H_3$,
  respectively) of the quasi-radial ($l=0$) modes for a sequence of
  rotating stars of same central density. The angular velocity $\Omega_K$
  at the mass-shedding limit is $0.8094 \times 10^4$ s$^{-1}$ for this
  sequence.}
\begin{tabular}{*{5}{c}}
\hline
 $\Omega$      &  $F$   & $H_1$ & $H_2$ & $H_3$ \\[0.5ex]
 $(10^4 s^-1)$ &  (kHz) & (kHz) & (kHz) & (kHz) \\[0.5ex]
\hline
\\[0.5ex]
0.0&    2.706&  4.547&  6.320&  8.153 \\[0.5ex]
0.218&  2.657&  4.467&  6.215&  8.005 \\[0.5ex]
0.306&  2.619&  4.409&  6.202&  8.005 \\[0.5ex]
0.371&  2.579&  4.385&  6.234&  8.096 \\[0.5ex]
0.399&  2.553&  4.377&  6.243&  8.098 \\[0.5ex]
0.423&  2.535&  4.371&  6.241&  8.134 \\[0.5ex]
0.445&  2.510&  4.362&  6.266&  8.171 \\[0.5ex]
0.465&  2.495&  4.356&  6.262&  8.171 \\[0.5ex]
0.482&  2.476&  4.366&  6.274&  8.197 \\[0.5ex]
0.498&  2.456&  4.357&  6.270&  8.130 \\[0.5ex]
0.511&  2.442&  4.350&  6.297&  8.030 \\[0.5ex]
0.522&  2.417&  4.337&  6.255&  7.987 \\[0.5ex]
\hline
\end{tabular}
\label{qr_tab}
\end{minipage}
\end{table}

As frequencies of quasi-radial modes have been computed previously by
Yoshida \& Eriguchi (2000), as an eigenvalue problem, we use those results 
to compare the values obtained with our code for a soft polytrope, with 
$\rho_{\mbox{c}}=8.1\times 10^{-4}$, $N=1.5$ and $K=4.349$.
We have compared the models corresponding to $\Omega=0, 7.1379\times 10^{-3}$
and $1.4094\times 10^{-2}$, which, in the notation of Yoshida \& Eriguchi
correspond to $f_{\mbox{rot}}= 0, 0.1415$ and $0.2794$, respectively.
The results of this comparison are presented in Table~\ref{YE}. We
note that the agreement is very good, especially for the most rapidly
rotating model, the differences always being below 2\%.  As there may
be some small differences in the construction of the equilibrium
model, the actual accuracy of our code is better than the relative
differences shown in Table~\ref{YE}. 

\begin{table}
\centering
\begin{minipage}{65mm}
\caption{Comparison of $l=0$ quasi-radial pulsation frequencies, obtained with the
  present nonlinear evolution code, to linear perturbation mode
  frequencies in the relativistic Cowling approximation (Yoshida \&
  Eriguchi 2000). The equilibrium model is a $N=1.5$, $K=4.349$
  relativistic polytrope with $\rho_{\mbox{c}}=8.1\times 10^{-4}$.}
\begin{tabular}{*{5}{c}}
\hline
$\Omega$ & Mode   & Y\&E & present & Difference
\\[0.5ex]
$(10^{-3})$ &        & (kHz)               & (kHz)      & (per cent)
\\[0.5ex]
\hline
0&              F&     1.674&  1.678&  0.3   \\[0.5ex]
          &    H1&     2.758&  2.807&  1.7   \\[0.5ex]
          &    H2&     3.793&  3.841&  1.3   \\[0.5ex]
\\[0.5ex]
$7.1379$& F&  1.646& 1.670& 1.5 \\[0.5ex]
                        & H1& 2.696& 2.735& 1.4 \\[0.5ex]
                        & H2& 3.728& 3.761& 0.9 \\[0.5ex]
\\[0.5ex]
$14.094$& F&  1.545& 1.553& 0.5 \\[0.5ex]
                        & H1& 2.572& 2.595& 0.9 \\[0.5ex]
                        & H2& 3.664& 3.642& 0.6 \\[0.5ex]
\hline
\end{tabular}
\label{YE}
\end{minipage}
\end{table}

\subsection{Non-radial modes}

\begin{figure}
\centerline{\psfig{file=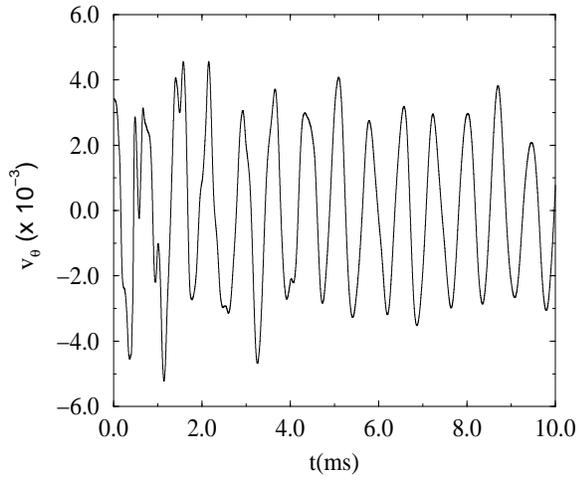,width=7.5cm,height=6.5cm}}
\caption{Time evolution of the polar velocity component for the fastest
rotator, $\Omega=0.522\times 10^4 s^{-1}$. An $l=1$ perturbation
has been applied to the equilibrium data.}
\label{l=1velz}
\end{figure}

\begin{figure}
\centerline{\psfig{file=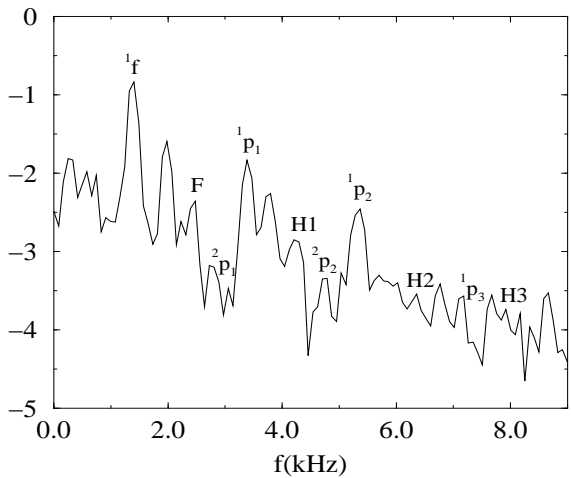,width=7.5cm,height=6.5cm}}
\caption{Fourier transform of the polar velocity evolution shown in 
Fig.~\ref{l=1velz}. The different axisymmetric, $l=1$ modes are conveniently 
labelled.}
\label{l=1velz_fft}
\end{figure}

In Figures~(\ref{l=1velz}-\ref{l=3velz_fft}) we plot the time
evolutions of the polar velocity component, along with the
corresponding Fourier transforms for non-radial $l=1,2$ and 3
perturbations, for the fastest rotating model in Table 1.  The time
evolutions shown are measured at half the star radius and
$\theta=2\pi/3$ for $l=1,3$, and $\theta= \pi/3$ for $l=2$. The above
time evolutions show the same qualitative behavior already described
for the $l=0$ modes. The lowest frequency (and dominant) $l=1$ mode we
excite in our time-evolutions (labeled $^1f$) has no nodes along the
radial direction and behaves like a fundamental mode. We point out
that the $l=1$ fundamental mode does not exist when one considers the
full set of equations (i.e. including the perturbation of the metric),
due to momentum conservation (it would correspond to a displacement of
the center of mass).  In the Cowling approximation, however, momentum
is not conserved, as the perturbation in the metric is neglected.  In
this approximation, only the fluid oscillates in a fixed background
metric and an oscillation of the center of mass is allowed, as the
fixed metric acts as a restoring force.

For $l=2$, a larger number of modes is excited by the initial
perturbation. At late times, the evolution is mainly a superposition of the
fundamental $^2f$-mode and the fundamental quasi-radial $F$-mode, as
it is evident from the amplitudes of the various modes in the
corresponding Fourier transform.

For $l=3$, a large number of modes is also excited by the initial
perturbation, as in the $l=2$ case. At late times, the evolution is
mainly a superposition of the fundamental $^3f$- and $^1f$-modes and
of the fundamental quasi-radial $F$-mode. The Fourier amplitude of the
$^1f$-mode is larger than the amplitude of the $^3f$-mode, which shows
that the eigenfunction of the $l=3$ modes near the mass-shedding limit
is significantly modified by rotation, so that the $l=3$ part of the
$^1f$-eigenfunction correlates better with the generic eigenfunction
we used to excite $l=3$ modes, than the $^3f$-eigenfunction itself.

\begin{figure}
\hspace{1.5cm}
\psfig{file=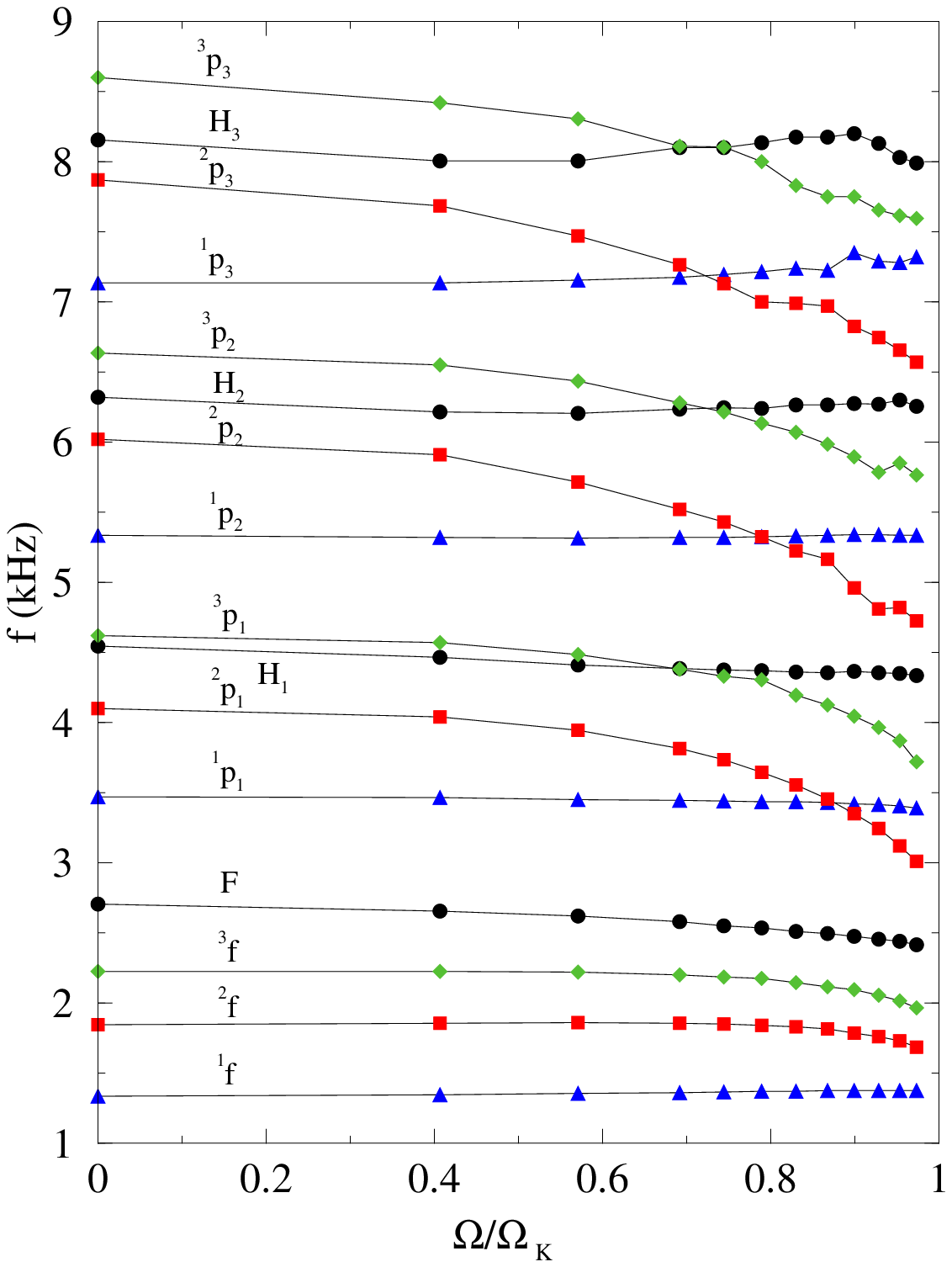,width=13cm,height=21cm}
\caption{Frequencies of the lowest three quasi-radial modes vs.
  the ratio of angular velocity of the star $\Omega$ to the angular
  velocity at the mass-shedding limit $\Omega_K$, for the sequence of
  rotating relativistic stars in Table \ref{initial_models}.}
\label{modes_fig}
\end{figure}

The frequencies of all identified $l=1,2$ and 3 modes are displayed in
detail in Tables~\ref{l1_tab}-\ref{l3_tab}. The modes are labeled as
$^lf$ for the fundamental modes and $^lp_n$ for the $p$-modes of order
$n$.  A plot of all mode-frequencies as a function of the rotation rate 
is shown in Figure~\ref{modes_fig}. At rotation rates below $\sim 50$\% 
of the mass-shedding limit, the frequencies of the lowest-order modes are 
not significantly affected by rotation. This is consistent with previous 
results in the slow-rotation approximation (Hartle \& Friedman 1975) and 
in the Newtonian limit. For larger rotation rates, however, the
$l=0$ and $l=1$ overtones have a tendency to increase in frequency
with rotation rate, while the $l=2$ and $l=3$ overtones have the
opposite tendency. As a result, several apparent mode crossings take
place between different modes. This has been observed before by
Clement (1986) for Newtonian axisymmetric modes and by Yoshida \&
Eriguchi (2000) for the relativistic quasi-radial modes. 

In the above studies, when one follows an eigenfrequency continuously
from the non-rotating limit to the large rotation rates, then, at
apparent mode crossings, the continuous lines corresponding to
different frequency sequences do not cross, which is normally called
``avoided crossing''.  Such avoided crossings can occur in two ways:
the eigenfunction along a continuous frequency sequence remains that
of the same mode (which is the usual type of avoided crossing in
non-rotating stars, see Unno et al 1989) or a different mode appears
in the same continuous frequency sequence after the avoided crossing.
The latter case is encountered in rotating stars, as in
Figure~\ref{modes_fig}. To distinguish the two cases, we prefer to use
the term ``apparent crossing'' for rotating stars (as was done in
Clement 1986), which refers to a mode-sequence, rather than a
frequency sequence. 

Finally, we note that in Clement (1981) it is claimed that the
axisymmetric $^2f$-mode lies on a continuous branch, near the
mass-shedding limit, for rapidly rotating Newtonian stars. In our
relativistic computation, no such behavior was found for any of the
modes studied. Within the numerical resolution employed all identified
modes were found to be discrete.

\subsection{Inertial modes}

\begin{figure}
\centerline{\psfig{file=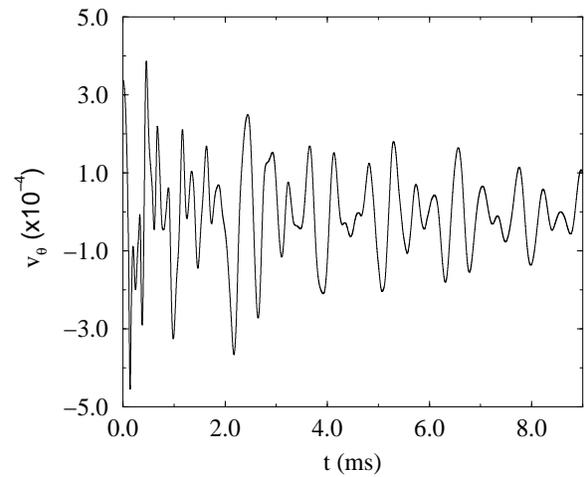,width=7.5cm,height=6.5cm}}
\caption{Same as Fig.~\ref{l=1velz} but showing the time evolution for the
$l=2$ perturbation.}
\label{l=2velz}
\end{figure}

\begin{figure}
\centerline{\psfig{file=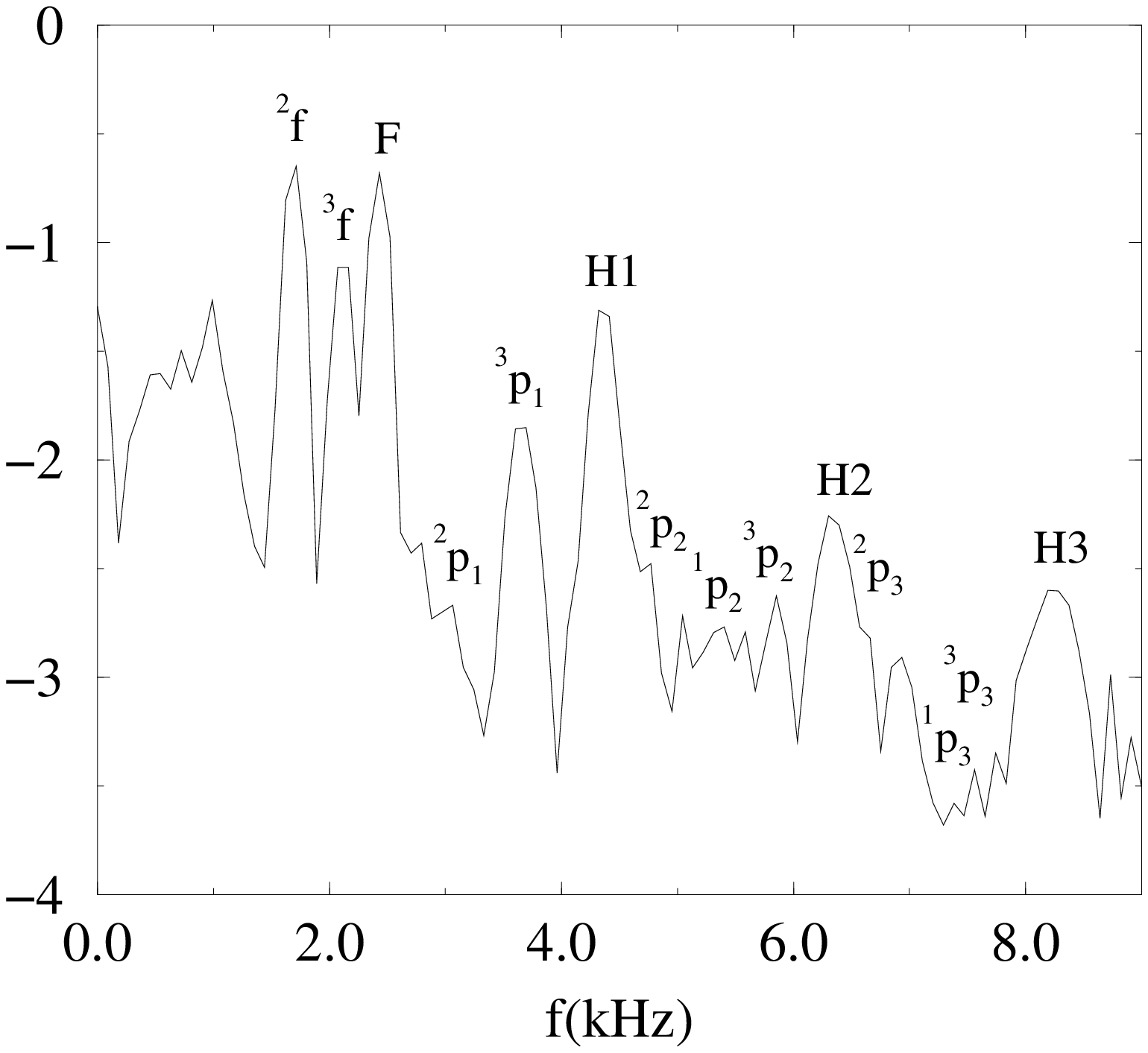,width=7.5cm,height=6.5cm}}
\caption{Same as Fig.~\ref{l=1velz_fft} but showing the frequencies of the
$l=2$ axisymmetric modes.}
\label{l=2velz_fft}
\end{figure}

\begin{figure}
\centerline{\psfig{file=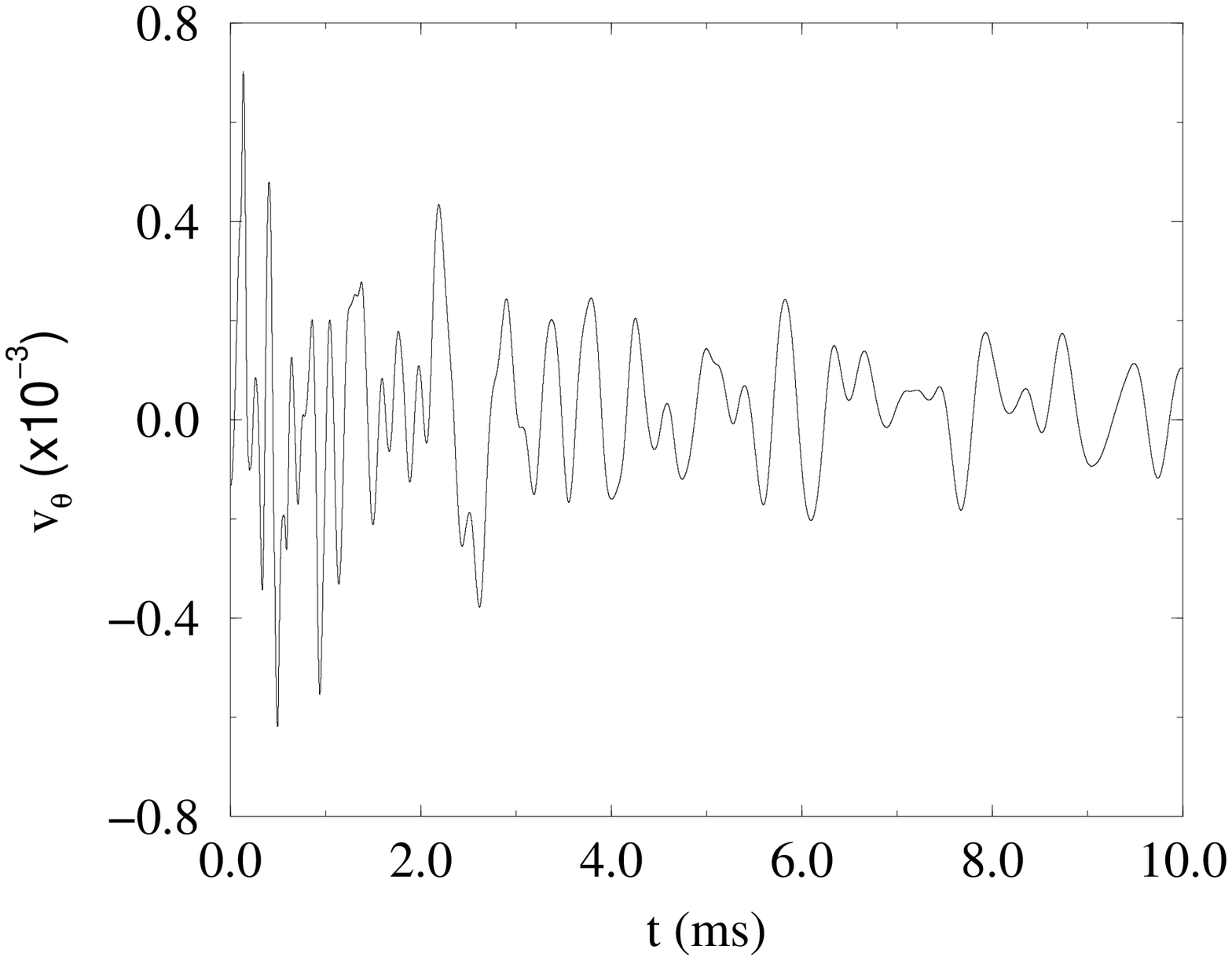,width=7.5cm,height=6.5cm}}
\caption{Same as Fig.~\ref{l=1velz} but showing the time evolution for the
$l=3$ perturbation.}
\label{l=3velz}
\end{figure}

\begin{figure}
\centerline{\psfig{file=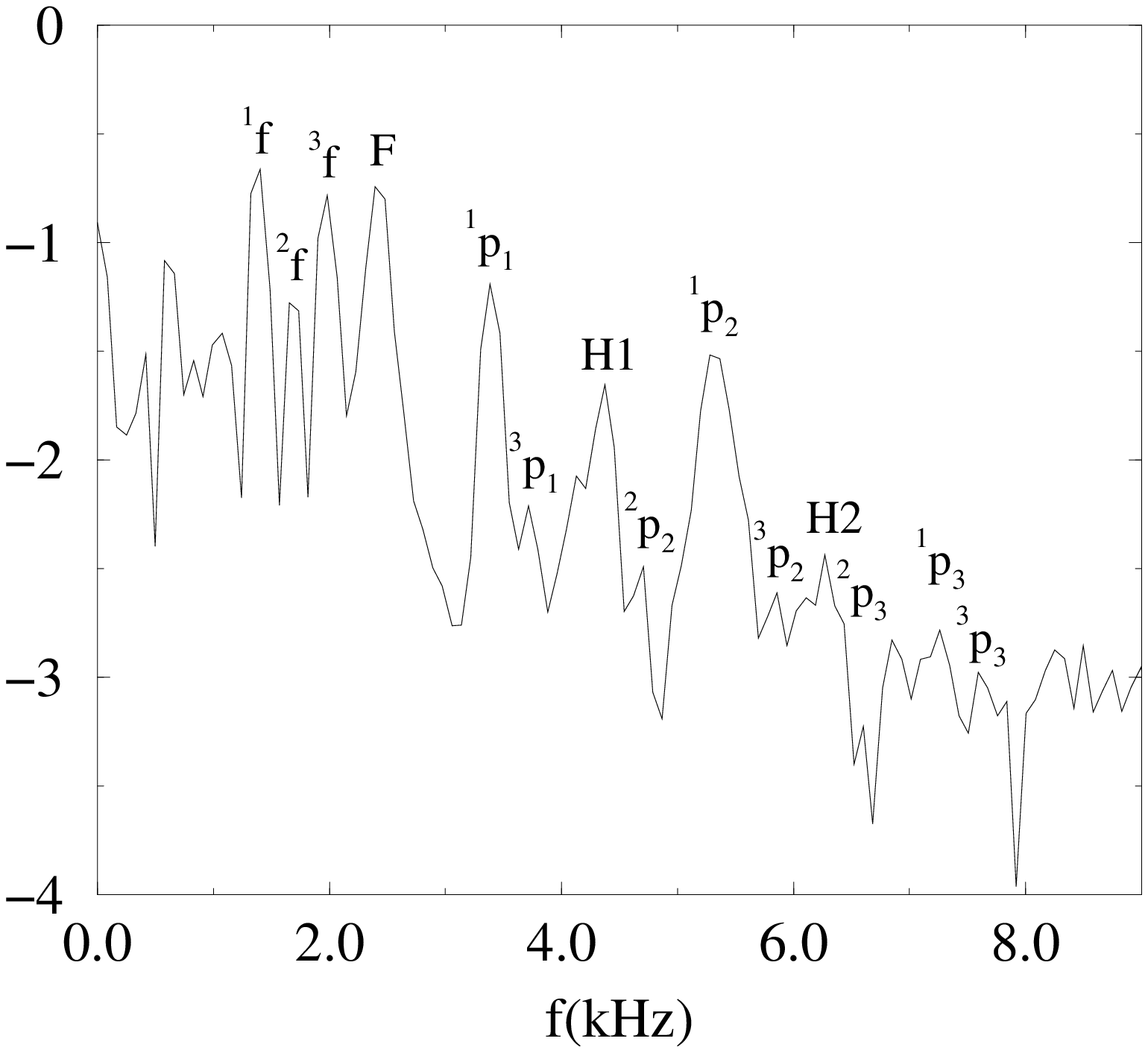,width=7.5cm,height=6.5cm}}
\caption{Same as Fig.~\ref{l=1velz_fft} but showing the frequencies of the
$l=3$ axisymmetric modes.}
\label{l=3velz_fft}
\end{figure}

Apart from the quasi-radial and $f$- and $p$-modes, a number of
axisymmetric inertial modes was also excited in our numerical
evolutions, which can be seen as low-frequency peaks in our Fourier
transforms (Figures~\ref{l=0velr_fft}-\ref{l=3velz_fft}).  These modes
exist in isentropic stars (such as those considered here) as a mixture
of axial $r$-modes and polar $g$-modes (see Lockitch \& Friedman 1999).
Nonaxisymmetric inertial modes have been computed as an eigenvalue
problem for slowly-rotating relativistic stars (Lockitch 1999,
Lockitch, Andersson \& Friedman 2000), but frequencies for axisymmetric
modes are not available yet. This makes their identification
difficult, as, in our simulations, many modes with similar frequencies
appear.  Furthermore, the spacing between these frequencies is of the
same order as one would expect the difference between the relativistic
frequencies and the Newtonian frequencies (computed in Lockitch \&
Friedman 1999) to be. Therefore, it would be too venturous, at this
point, to attempt an identification with specific normal modes,
without prior knowledge of some of these frequencies in relativity, at
least for slow-rotation.

\begin{table}
\centering
\begin{minipage}{63mm}
\caption{$f$-mode and first three $p$-modes corresponding to {\it $l=1$}.
Further details in Tables \ref{initial_models} and \ref{qr_tab}.}
\begin{tabular}{*{5}{c}}
\hline
$\Omega$      &  $^1f$  & $^1p_1$  & $^1p_2$  & $^1p_3$  \\[0.5ex]
$(10^4 s^-1)$ &  (kHz)  & (kHz)    & (kHz)    & (kHz)    \\[0.5ex]
\hline
\\[0.5ex]
0.0&    1.335&  3.473&  5.335&  7.136 \\[0.5ex]
0.218&  1.349&  3.464&  5.318&  7.134 \\[0.5ex]
0.306&  1.356&  3.453&  5.317&  7.152 \\[0.5ex]
0.371&  1.364&  3.446&  5.320&  7.172 \\[0.5ex]
0.399&  1.369&  3.442&  5.322&  7.193 \\[0.5ex]
0.423&  1.371&  3.438&  5.325&  7.214 \\[0.5ex]
0.445&  1.373&  3.434&  5.328&  7.238 \\[0.5ex]
0.465&  1.375&  3.429&  5.333&  7.223 \\[0.5ex]
0.482&  1.376&  3.422&  5.339&  7.349 \\[0.5ex]
0.498&  1.376&  3.417&  5.340&  7.288 \\[0.5ex]
0.511&  1.375&  3.407&  5.337&  7.281 \\[0.5ex]
0.522&  1.375&  3.393&  5.335&  7.318 \\[0.5ex]
\hline
\end{tabular}
\label{l1_tab}
\end{minipage}
\end{table}

\begin{table}
\centering
\begin{minipage}{63mm}
\caption{$f$-mode and first three $p$-modes corresponding to {\it $l=2$}.
Further details in Tables \ref{initial_models} and \ref{qr_tab}.}
\begin{tabular}{*{5}{c}}
\hline
$\Omega$      &  $^2f$  & $^2p_1$  & $^2p_2$  & $^2p_3$  \\[0.5ex]
$(10^4 s^-1)$ &  (kHz)  & (kHz)    & (kHz)    & (kHz)    \\[0.5ex]
\hline
\\[0.5ex]
0.0&    1.846&  4.100&  6.019&  7.867 \\[0.5ex]
0.218&  1.855&  4.040&  5.910&  7.684 \\[0.5ex]
0.306&  1.860&  3.944&  5.716&  7.471 \\[0.5ex]
0.371&  1.857&  3.814&  5.521&  7.264 \\[0.5ex]
0.399&  1.851&  3.734&  5.431&  7.130 \\[0.5ex]
0.423&  1.844&  3.645&  5.325&  7.000 \\[0.5ex]
0.445&  1.832&  3.554&  5.226&  6.989 \\[0.5ex]
0.465&  1.815&  3.456&  5.164&  6.970 \\[0.5ex]
0.482&  1.787&  3.352&  4.962&  6.823 \\[0.5ex]
0.498&  1.762&  3.244&  4.810&  6.746 \\[0.5ex]
0.511&  1.733&  3.120&  4.822&  6.653 \\[0.5ex]
0.522&  1.686&  3.010&  4.726&  6.571 \\[0.5ex]
\hline
\end{tabular}
\label{l2_tab}
\end{minipage}
\end{table}

\begin{table}
\centering
\begin{minipage}{63mm}
\caption{$f$-mode and first three $p$-modes corresponding to {\it $l=3$}.
Further details in Tables \ref{initial_models} and \ref{qr_tab}.}
\begin{tabular}{*{5}{c}}
\hline
$\Omega$      &  $^3f$  & $^3p_1$  & $^3p_2$  & $^3p_3$  \\[0.5ex]
$(10^4 s^-1)$ &  (kHz)  & (kHz)    & (kHz)    & (kHz)    \\[0.5ex]
\hline
\\[0.5ex]
0.0&    2.228&  4.622& 6.635    & 8.600 \\[0.5ex]
0.218&  2.228&  4.570& 6.550    & 8.418 \\[0.5ex]
0.306&  2.221&  4.485& 6.433    & 8.304 \\[0.5ex]
0.371&  2.199&  4.380& 6.280    & 8.109 \\[0.5ex]
0.399&  2.186&  4.330& 6.214    & 8.105 \\[0.5ex]
0.423&  2.177&  4.303& 6.135    & 8.000 \\[0.5ex]
0.445&  2.148&  4.194& 6.067    & 7.831 \\[0.5ex]
0.465&  2.118&  4.124& 5.987    & 7.751 \\[0.5ex]
0.482&  2.094&  4.044& 5.895    & 7.751 \\[0.5ex]
0.498&  2.055&  3.964& 5.784    & 7.656 \\[0.5ex]
0.511&  2.017&  3.870& 5.852    & 7.612 \\[0.5ex]
0.522&  1.965&  3.720& 5.767    & 7.593 \\[0.5ex]
\hline
\end{tabular}
\label{l3_tab}
\end{minipage}
\end{table}

\section{Summary and Outlook}                                           %
\label{summary}                                                         %

We have presented a comprehensive study of all low-order axisymmetric
modes of uniformly and rapidly rotating relativistic stars in the
Cowling approximation.  This investigation has been carried out by
numerically evolving initial perturbed equilibrium configurations with
an axisymmetric, nonlinear, relativistic hydrodynamics code. The
simulations were performed using a high-resolution shock-capturing
finite-difference scheme accurate enough to maintain the initial
rotation law for a sufficient number of rotational periods.

Through Fourier transforms of the time evolution of selected fluid
variables we computed the frequencies of non-radial, axisymmetric modes
(with angular momentum indices $l=0,1,2$ and $3$) of rapidly rotating
stars. Therefore, we have extended previous results by Yoshida \&
Eriguchi (2000) which were mainly restricted to quasi-radial modes. We
have presented results for a complete sequence of rotating stars,
ranging from the non-rotating case to rapid rotation near the
mass-shedding limit. Apparent crossings between different modes appear
for rapidly rotating stars, as a result of the different influence
of rotation on quasi-radial and $l=1$ modes than on modes with $l\geq 2$.
This different behavior may be related to the fact that the rotational
deformation of the equilibrium star appears first as an $l=2$ term.
Several axisymmetric inertial modes were also excited in our simulations.
However, a definitive identification of the observed frequency peaks
with specific modes will only be possible when mode-frequencies on
the slow-rotation approximation be computed as an eigenvalue problem.
Alternatively, a determination of mode-eigenfunctions in our simulations
(and a comparison to mode-eigenfunctions in the Newtonian limit) may
also allow the identification of such inertial modes. 

In following work we plan to study axisymmetric modes of
differentially rotating stars with realistic equations of state.
Moreover, the implementation of the hydrodynamic equations and
numerical techniques employed in the present work has been recently
extended (Dimmelmeier et al 2000) to allow for gravitational field
dynamics through the so-called {\it conformally flat metric} approach
(Wilson et al 1996). Studies of fully coupled evolutions with such a
code, in the context of pulsations of rotating relativistic stars,
will be presented elsewhere.

\section*{Acknowledgements}                                             %

We thank John Friedman, Kostas Kokkotas and Ewald M{\"u}ller for helpful
discussions and comments on the manuscript.  All computations have
been performed on a NEC SX-5/3C Supercomputer at the Rechenzentrum
Garching. A.G. thanks the Max-Planck-Institut f{\"u}r Gravitationsphysik,
Golm and the Max-Planck-Institut f{\"u}r Astrophysik, Garching, for
supporting a visit during which this collaboration was initiated.
This research was supported in part by the European Union grant
HPRN-CT-2000-00137.


\newpage

\onecolumn

\twocolumn
\end{document}